\documentclass[epsfig,12pt]{article}
\usepackage{epsfig}
\usepackage{graphicx}

\usepackage{latexsym}
\usepackage{amsmath}
\usepackage{amssymb}
\usepackage{relsize}
\usepackage{geometry}
\def\beq{\begin{equation}}
\def\eeq{\end{equation}}
\def\beqn{\begin{eqnarray}}
\def\eeqn{\end{eqnarray}}

\newcommand{\none}{${\mathcal N}=1\,$}
\newcommand{\ntt}{${\mathcal N}=(2,2)\,$}
\newcommand{\nzt}{${\mathcal N}=(0,2)\,$}
\newcommand{\cpn}{CP$(N-1)\,$}

\newcommand{\cell}{{\mathcal L}}

\newcommand{\pt}{\partial}

\newcommand{\cf}{${\mathcal F}$}
\newcommand{\cfe}{{\mathcal F}}
\newcommand{\cse}{{\mathcal S}}
\newcommand{\cbe}{{\mathcal B}}
\newcommand{\cwe}{{\mathcal W}}

\newcommand{\gsim}{\lower.7ex\hbox{$
\;\stackrel{\textstyle>}{\sim}\;$}}
\newcommand{\lsim}{\lower.7ex\hbox{$
\;\stackrel{\textstyle<}{\sim}\;$}}

\begin{document}


\begin{titlepage}

\begin{flushright}
FTPI-MINN-09/48, UMN-TH-2830/09\\
\end{flushright}

\vspace{1cm}

\begin{center}
{  \Large \bf   \boldmath{\nzt} Deformation of the \boldmath{\ntt} \\[2mm]
Wess-Zumino Model
in Two Dimensions}
\end{center}

\vspace{1mm}

\begin{center}

 {\large
 \bf   M.~Shifman$^{\,a}$ and \bf A.~Yung$^{\,\,a,b}$}
\end {center}

\begin{center}


$^a${\it  William I. Fine Theoretical Physics Institute,
University of Minnesota,
Minneapolis, MN 55455, USA}\\
$^{b}${\it Petersburg Nuclear Physics Institute, Gatchina, St. Petersburg
188300, Russia
}
\end{center}

\vspace{2cm}

\begin{center}
{\large\bf Abstract}
\end{center}
We construct a simple  \nzt deformation of the two-dimensional 
Wess--Zumino model. In addition to superpotential, it includes a ``twisted"
superpotential. Supersymmetry may or may not be spontaneously broken
at the classical level. In the latter case an extra right-handed fermion field
$\zeta_R$ involved in the \nzt deformation plays the role of Goldstino.

\vspace{2cm}

\end{titlepage}

\newpage



Recently it was found \cite {EdTo,SY1} that non-Abelian string solitons in certain \none bulk gauge theories
are described on the world sheet by \nzt deformations of the \cpn models. 
This finding raised interest to \nzt deformations of two-dimensional \ntt models in general.
Here we will consider \mbox{\nzt} deformations of the Wess--Zumino model \cite{WeZu}.
General elements of \nzt de\-formations were worked out by Witten \cite{W93,W95}.
A broad class of
the $(0,2)$ Landau--Ginzburg models were analyzed, from various perspectives, in \cite{D1,D2,D3}.
The prime interest of these studies was the flow of the  (0,2) Landau-Ginzburg models  to
non-trivial (0,2) superconformal field theories \cite{D1,D2}, and  \nzt analogs of the topological rings
in the \ntt theories \cite{D3}.

Here we will consider the \nzt deformation of the Wess--Zumino model 
with the emphasis on an aspect which will be thoroughly studied in a subsequent publication
\cite{D4}, namely, spontaneous breaking/nonbreaking of supersymmetry.
Related issues of interest are (i) a nonrenormalization theorem; (ii)
BPS saturation of possible kinks.
We use a  formalism which is simple enough and is adequate to the problem.
It parallelizes the formalism exploited in \cite{SY1} to construct the heterotic \cpn model,
see also \cite{BSY3}. For simplicity we  consider only the simplest version of the Wess--Zumino model,
in which interactions come only from the potential term. Generalizations are straightforward,
see also \cite{D1,D2}.

We find that, even though the \nzt supersymmetry is implemented
at the Lagrangian level, generically supersymmetry is spontaneously broken
at the tree level. One can fine-tune a free parameter of the model  in such a way that
it stays unbroken at the tree level, and then (presumably) to any finite order in perturbation
theory.

Two space-time coordinates are
\beq
x^\mu =\{t,\,z\}\,,\qquad \mu =0,1.
\label{1}
\eeq
The \ntt superspace is spanned by\,\footnote{The gamma-matrices  are chosen as 
$
\gamma^{0}=\gamma^t=\sigma_2\,,\qquad \gamma^{1}=\gamma^z = i\sigma_1\,,\qquad \gamma_{5} 
\equiv\gamma^0\gamma^1 = \sigma_3
$. Moreover, $\bar \theta=\theta^{\dagger}\gamma^{0}$. With these definitions,
the $\alpha = 1$ spinor component is right-handed while $\alpha = 2\,$ left-handed.
}
\beq
\{x^\mu,\,\theta_\alpha,\,\bar\theta_\beta\}\,,\qquad \alpha ,\,\beta =1,2.
\label{2}
\eeq
In addition to the standard chiral superfields $\Phi^a$ of the conventional 
Wess--Zumino model, we will introduce \nzt suerfields
\beqn
{\mathcal B} &=& \left\{\,\zeta_R (x^\mu +i\bar\theta\gamma^\mu\theta) +\sqrt{2}\theta_R{\mathcal F}
\right\} \theta_L^\dagger\,,
\nonumber\\[2mm]
{\mathcal B}^\dagger &=&\theta_L \left\{\,\zeta_R^\dagger (x^\mu - i\bar\theta\gamma^\mu\theta) +\sqrt{2}\theta_R^\dagger {\mathcal F}^\dagger
\right\} \,.
\label{AAA}
\eeqn
Since $\theta_L$ and $\theta_L^\dagger$ enter in Eq.~(\ref{AAA})  explicitly,
${\mathcal B}$ and ${\mathcal B}^\dagger$ are {\em not} superfields with regards
to the supertransformations with parameters $\epsilon_L,\, \epsilon_L^\dagger$.
These supertransformations are absent in the heterotic model.
Only those survive which are associated with  $\epsilon_R,\, \epsilon_R^\dagger$. 
Note that ${\mathcal B}$ and ${\mathcal B}^\dagger $
are superfields with regards to the shifts with $\epsilon_R,\, \epsilon_R^\dagger$. 
As usual, we will  introduce a shorthand for the chiral coordinate
\beq
\tilde{x}^\mu =x^\mu + i\bar\theta\gamma^\mu\theta\,.
\label{3}
\eeq
Then the transformation laws with the parameters $\epsilon_R,\, \epsilon_R^\dagger$
are as follows (we set $\epsilon_L = \epsilon_L^\dagger =0$):
\beq
\delta\theta_R =\epsilon_R\,,\quad \delta\theta_R^\dagger =\epsilon_R^\dagger
\,,\quad \delta\tilde{x}^0 = 2i \epsilon_R^\dagger\theta_R
\,,\quad \delta\tilde{x}^1 = 2i \epsilon_R^\dagger\theta_R\,.
\label{4}
\eeq
With respect to such supertransformations, ${\mathcal B}$ and ${\mathcal B}^\dagger$ are  superfields.
Indeed,
\beq
\delta\zeta_R = \sqrt{2}\, {\mathcal F} \,\epsilon_R\,,\quad  \delta {\mathcal F} =\sqrt{2}\,i\left(\partial_L
\zeta_R\right) \epsilon_R^\dagger\,,
\label{5}
\eeq
plus Hermitian conjugate transformations. 

Thus, the boson sector of the deformed model coincides with that of the conventional
Wess--Zumino model, while the fermion sector is expanded. In addition to the fermion fields $\psi_{R,L}^a$
of the Wess--Zumino model it includes a right-handed fermion field $\zeta_R$.

The \nzt action can be written as
\beqn
S 
&=&
 \int d^2x\,\left\{ d^4\theta \, \Phi^{a\,\dagger}\Phi^a +\left[d^2\theta\,  {\mathcal W}(\Phi^a ) +{\rm H.c.}\right]
+\Delta \cell_h \right\}
\,,
\nonumber\\[3mm] 
\Delta\cell_h
&=&
\left\{\sqrt{2}\kappa
\int d\theta_L^\dagger\, d\theta_R\,  
\, {\mathcal B}\,  +{\rm H.c.}\right\}
\nonumber\\[3mm]
&-&
2
\int  d^4\theta \,{\mathcal B}^\dagger{\mathcal B} + 2\left\{
\int d\theta_L^\dagger\, d\theta_R\, \theta_L\, d\theta_L
\, {\mathcal B}\, {\mathcal S}(\Phi^a) +{\rm H.c.}\right\}\,,
\label{7}
\eeqn
where the second term presents the heterotic deformation, ${\mathcal W}$ is the superpotential,
while ${\mathcal S}(\Phi^a) $ is a function of the chiral superfield $\Phi$ which couples the heterotic sector
to the conventional Wess--Zumino model. (Some generalizations will be considered later.)
Let us call it $h$-superpotential. 
Moreover, $\kappa$ is a constant of dimension of mass.
Note that both, superpotential and $h$-superpotential have dimensions of mass too.
The terms containing ${\mathcal B}$
and given by integrals over a reduced superspace
 will be referred to as $h$ terms.
 Adding a constant  to the $h$-superpotential is equivalent
to shifting $\kappa$ since they enter only in the combination
$\cse (\phi^a) +\kappa$, see Eq.~(\ref{9}).
One can use this freedom to fix the value of ${\mathcal S}$ at some given point, without
loss of generality. 

In components
\beqn
\cell 
&=&
 \pt_\mu \phi^{a\dagger} \pt^\mu \phi^a 
+\bar\psi^a\gamma^\mu\, i\,\pt_\mu\psi^a + F^{a\,\dagger}\,F^a
\nonumber\\[3mm]
&+&
\left\{F^a\,\pt_a{\mathcal W} -\left(\pt_a\pt_b{\mathcal W}\right)
\left(\psi_L^a\psi_R^b \right) +{\rm H.c.}
\right\} +\Delta\cell_h
\label{8}
\eeqn
where
\beq
\Delta\cell_h =
\zeta_R^\dagger\,  i\pt_L\zeta_R +\cfe^\dagger\cfe
+
\left\{ \cfe\left[ \cse (\phi^a) +\kappa\right]  -\zeta_R \psi^a_L\,\pt_a\cse +{\rm H.c.} \right\},
\label{9}
\eeq
and
\beq
\pt_L =\pt_t +\pt_z\,,\qquad \pt_R =\pt_t -\pt_z\,.
\label{9prim}
\eeq
The auxiliary superfield \cf, as usual, can be eliminated via equations of motion,
\beq
\cfe^\dagger = - \left[ \cse (\phi^a) +\kappa\right].
\label{10}
\eeq
Then the bosonic part of $\cell_h$ takes the form
\beq
\cell_{h,{\rm bos}} =- \left| \cse (\phi^a) +\kappa\right|^2\,.
\label{11}
\eeq
Adding the Wess--Zumino part we obtain the scalar potential,
\beq
V = \left| \pt_a{\mathcal W} \right|^2 +\left| \cse (\phi^a) +\kappa\right|^2\,.
\label{12}
\eeq
Supersymmetric vacua exist (at the classical level) provided that the set of equations
\beq
\pt_a{\mathcal W}=0 \qquad ({\rm all}\,\,\, a), \qquad \cse (\phi^a) +\kappa =0
\label{13}
\eeq
are satisfied
at one or more critical points $\phi_*$. Considering $\kappa$ as a free parameter one 
can always fine-tune it in such a way
that at least one vacuum (a solution $\pt_a{\mathcal W}(\phi_*)=0$) will be classically supersymmtric.

It is instructive to derive conserved supercurrents. If in the undistorted model with \ntt supersymmetry
we had four conserved supercurrents, $J^\mu_L ,\,\, J^\mu_R$ and 
their complex conjugated, now we expect only two of those to survive. The conserved components are
\beqn
J^\mu_L
& =&
 \sqrt{2}\left\{ i\, \nu^\mu\,  \psi^\dagger_R\, F + i\, \nu^\mu\,  \zeta^\dagger_R\, \cfe
+\bar\nu^\mu\,\psi_L \pt_L\phi^\dagger
\right\}
\nonumber\\[3mm]
\left(J^\mu_L\right)^\dagger
& =&
 \sqrt{2}\left\{ -i\, \nu^\mu\,  \psi_R\, F^\dagger - i\, \nu^\mu\,  \zeta_R\, \cfe^\dagger
+\bar\nu^\mu\,\psi_L^\dagger \pt_L\phi
\right\}
\label{14} 
\eeqn
where summation over $a$ is implicit, and we defined two conjugated 2-vectors,
\beq
\nu^\mu = \{1,\,\,1\}\,,\qquad \bar\nu^\mu = \{1,\,\,-1\}\,.
\label{15} 
\eeq
The corresponding superalgebra is as follows:
\beq
\{Q^\dagger_L\,\,Q_L\} = 2\left(H - P^z\right).
\label{16} 
\eeq
From Eq.~(\ref{16}) it is obvious that massless right-movers  can (an do) form short (single-state) ``multiplets."

In the \ntt Wess--Zumino model ($\cbe =0$) there is a relation for the dilatation operator
\beq
\left(\gamma_\mu\, J^\mu  \right)_{L,R} = i\,2\sqrt{2} \,F\left(\psi^\dagger\right)_{L,R}\,.
\label{17} 
\eeq
In the \nzt-deformed model the analog of this relation is
\beq
\bar\nu_\mu\, J^\mu_L  =  i\,2\sqrt{2} \left( F\,\psi^\dagger_R + \cfe \zeta_R^\dagger\right).
\label{18} 
\eeq

\vspace{3mm}

\underline{{\em Some generalizations}}

\vspace{1mm}

In addition to (\ref{9}), one can couple the $\cbe$ field
to other fields  through a number of extra terms, for instance,
\beq
\int d^4\theta\, \cbe^\dagger\cbe f(\Phi\Phi^\dagger)\,\,\,\mbox{or}\,\,\, \int d^4\theta\, \cbe \tilde f(\Phi\Phi^\dagger) +{\rm H.c.}
\label{20}
\eeq
The first term gives, in particular,  a coupling of the $\zeta$ kinetic term with
the $\phi,\,\phi^\dagger$ fields. As was mentioned, such interactions will not be considered for the time being.
The second term was considered in \cite{SY1}. It generates the
$\zeta_R\,\pt_L\phi^\dagger\, \psi_R$ interaction and an additional bifermion term $\psi^\dagger_L\,\psi_R$
in (\ref{10}), as well as $\zeta^\dagger_R\psi_L$ in $F$, resulting in   four-fermion interactions in the Lagrangian.

\vspace{3mm}

\underline{{\em Nonrenormalization of $h$ terms}}

\vspace{1mm}

As well-known, $F$ terms 
in the effective Lagrangian in the \ntt theory are protected from renormalizations
by nonrenormalization theorems \cite{nonrenormth,natiseib}. Thus, the superpotential term, being an integral over a reduced superspace,
is unaffected by loops. Since the $h$-terms are also given by  integrals over a reduced superspace,
similar theorems can be establishes in the \nzt models for these terms.
An appropriate choice of the background field in this case is
\beqn
 \Phi^\dagger_{\rm b} 
 &=&
  0\, , \qquad\Phi_{\rm b} = C_1 +C_2^\alpha\theta_\alpha 
+C_3\theta^2\, ,
\nonumber\\[2mm]
\cbe^\dagger_{\rm b} 
&=& 0
\, , \qquad
\cbe_{\rm b}  = C_4 \theta_L^\dagger+  C_5 \theta_R\theta_L^\dagger\,,
\label{cbf}
\eeqn
where 
the subscript b mark the background fields, and
$C_{1,2,3,4,5}$ are  $c$-numerical {\em constants}. This choice assumes that $\Phi$ and
$\Phi^\dagger$ are treated as independent variables, not connected by 
complex conjugation, and so are $\cbe$ and $\cbe^\dagger$ (i.e. we keep in mind a kind of analytic 
continuation).
The $x$ independent fields (\ref{cbf}) are invariant under the 
action of
$Q_L^\dagger$.
Next, to do the calculation of the effective action
we decompose the superfields
\beqn
\Phi  &=&
\Phi_{\rm b} +\Phi_{\rm qu}\, , \,\,\,\,   \Phi^\dagger =  \Phi_{\rm b} 
+ \Phi_{\rm qu}^\dagger\, ,
\nonumber\\[2mm]
 \cbe
 &=&
 \cbe_{\rm b} +\Phi_{\rm qu}\, , \,\,\,\,   \cbe^\dagger =  \cbe_{\rm b} 
+ \cbe_{\rm qu}^\dagger\, ,
\eeqn
where the subscript qu denotes the quantum part of the superfield, 
expand 
the action in $\Phi_{\rm qu}, \cbe_{\rm qu}$ dropping the linear 
terms, and 
treat the remainder as the action for the quantum fields. We, then, 
integrate the quantum fields over, order by order, keeping the background 
field fixed. The crucial point is that in the given background field 
(a) the terms containing  ${\mathcal W}$ and $\cbe$, {\em without}
${\mathcal W}^\dagger$ and $\cbe^\dagger$, 
do
{\em not} vanish, and (b) there exists an exact supersymmetry under 
$Q_L^\dagger$-generated supertransformations.

 After substituting in loops   Green's 
functions in the given background and integrating over all vertices except the
first one we arrive at an expression of the type
\begin{equation}
\int   d \theta_R^\dagger \times \,\left( {\rm a} \,\,  \theta_R^\dagger
\,\,\mbox{independent 
function}\right) = 0\, .
\label{saclaypp}
\end{equation}
The $\theta_R^\dagger$ independence follows from the fact that our 
superspace is homogeneous in the $\theta_R^\dagger$ direction even in the presence of 
the background field (\ref{cbf}). This completes the proof   of 
nonrenormalization of $F$ and $h$ terms in the \nzt theory. 

A subtle point here is that this proof tacitly assumes the absence of infrared singularities.
Thus, it is certainly valid for the Wilsonean effective action \cite{SVWils}.
If infrared contributions are included (i.e. the generator of 1-particle irreducible amplitudes
is studied) the question should be investigated on a case by case basis.

\vspace{3mm}

\underline{{\em Kinks}}

\vspace{1mm}

\ntt models with two or more supersymmetric vacua (two or more zeros of $\pt_a\cwe$)
support interpolating kink solutions which, typically, are 1/2 BPS saturated, i.e. preserve two out of four
supersymmetries of the \ntt model under consideration. Adding an \nzt deformation
can destroy vacuum degeneracy and, thus, eliminate kinks altogether.
Even if we choose a deformatioin of a special form which does not break \nzt supersymmetry
in two or more vacua, BPS-saturated kinks do not exist in such theories.
This is readily seen through the Bogomoln'yi completion \cite {bogo}. With the heterotic deformation switched on
the Bogomol'nyi completion of the bosonic part of the energy functional is impossible, generally speaking.

\section*{Acknowledgments}


The work of MS was supported in part by DOE grant DE-FG02-94ER408. 
The work of AY was  supported 
by  FTPI, University of Minnesota, 
by RFBR Grant No. 09-02-00457a 
and by Russian State Grant for 
Scientific Schools RSGSS-11242003.2.

\end{document}